\documentclass[3p,twocolumn]{elsarticle}

\usepackage{amssymb}
\usepackage{amsmath}
\usepackage{graphics}
\usepackage{epsfig}
\usepackage[dvips]{color}
\usepackage{bm}
\usepackage{color}

\newcommand{\be}{\begin{equation}}

\newcommand{\ee}{\end{equation}}
\newcommand{\bea}{\begin{eqnarray}}
\newcommand{\eea}{\end{eqnarray}}
\newcommand{\beas}{\begin{eqnarray*}}
\newcommand{\eeas}{\end{eqnarray*}}

\newcommand{\nn}{\nonumber\\}

\newcommand{\me}{\mathrm{e}} 
\newcommand{\mi}{\mathrm{i}} 
\newcommand{\df}[1]{\ensuremath{\frac{\mathrm{d}^{3}#1}{(2\pi)^{3}}}\,}

\begin{document}

\title{The effective QCD phase diagram and the critical end point}

\author[icm,uct]{Alejandro Ayala}
\author[ifm]{Adnan Bashir}
\author[ifm]{J.J. Cobos-Mart\1nez}
\author[ifm]{Sa\'ul Hern\'andez-Ortiz}
\author[ifm]{Alfredo Raya}

\address[icm]{Instituto de Ciencias
  Nucleares, Universidad Nacional Aut\'onoma de M\'exico, Apartado
  Postal 70-543, M\'exico Distrito Federal 04510,
  M\'exico.}

\address[uct]{Centre for Theoretical and Mathematical Physics, and Department 
of Physics, University of Cape Town, Rondebosch 7700, South Africa.}

\address[ifm]{Instituto de F\1sica y Matem\'aticas, Universidad Michoacana de 
San Nicol\'as de Hidalgo, Edificio C-3, Ciudad Universitaria, Morelia, 
Michoac\'an 58040, M\'exico.}


\begin{abstract}

We study the QCD phase diagram on the temperature $T$ and
quark chemical potential $\mu$ plane, modeling the strong interactions
with the linear sigma model coupled to quarks. The phase
transition line is found from the effective potential at finite
$T$ and $\mu$ taking into account the plasma screening
effects. We find the location of the critical end point (CEP) to
be
$(\mu^{\mbox{\tiny{CEP}}}/T_c,T^{\mbox{\tiny{CEP}}}/T_c)\sim(1.2,0.8)$,
where $T_c$ is the (pseudo)critical temperature for the crossover
phase transition at vanishing $\mu$. This location lies within the
region found by lattice inspired calculations. The results show
that in the linear sigma model, the CEP's location in the phase
diagram is expectedly determined solely through chiral symmetry
breaking. The same is likely to be true for all other models which
do not exhibit confinement, provided the proper treatment of the
plasma infrared properties for the description of chiral symmetry
restoration is implemented. Similarly, we also expect these
corrections to be substantially relevant in the QCD phase diagram.

\end{abstract}


\maketitle

The different phases in which matter, made up of quarks and
gluons, arranges itself depends, as for any other substance, on
the temperature and density, or equivalently, on the temperature
and chemical potentials. Under the assumptions of beta decay
equilibrium and charge neutrality, the representation of the QCD
phase diagram is two dimensional. This is customary plotted with the
light-quark chemical potential $\mu$ as the horizontal variable
and the temperature $T$ as the vertical one.  $\mu$ is related to
the baryon chemical potential $\mu_B$ by $\mu_B=3\mu$.

Most of our knowledge of the phase diagram is restricted to the
$\mu=0$ axis. The phase diagram is, by and large, unknown. For
physical quark masses and $\mu=0$, lattice calculations have
shown~\cite{Aoki} that the change from the low temperature phase,
where the degrees of freedom are hadrons, to the high temperature
phase described by quarks and gluons, is an analytic crossover.
The phase transition has a dual nature: on the one hand the
color-singlet hadrons break up leading to deconfined quarks and
gluons; this is dubbed as the {\em deconfinement phase
transition}. On the other hand, the dynamically generated
component of quark masses within hadrons vanishes; this is
referred to as {\em chiral symmetry restoration}.

Lattice calculations have provided values for the crossover
(pseudo)critical temperature $T_c$ for $\mu=0$ and 2+1 quark
flavors using different types of improved rooted staggered
fermions~\cite{Levkova}. The MILC collaboration~\cite{MILC}
obtained $T_c=169(12)(4)$ MeV. The RBC-Bielefeld
collaboration~\cite{BNL-RBC-Bielefeld} reported $T_c=192(7)(4)$
MeV. The Wuppertal-Budapest
collaboration~\cite{Wuppertal-Budapest} has consistently obtained
smaller values, the latest being $T_c=147(2)(3)$ MeV. The HotQCD
collaboration has computed $T_c=154(9)$ MeV~\cite{HotQCD} and 
more recently $T_c=155(1)(8)$ MeV~\cite{HotQCD2}. The
differences could perhaps be attributed to different lattice spacings.

The picture presented by lattice QCD for $T\geq 0$, $\mu=0$ cannot
be easily extended to the case $\mu\neq 0$, the reason being that
standard Monte Carlo simulations can only be applied to the case
where either $\mu=0$ or is purely imaginary. Simulations with
$\mu\neq 0$ are hindered by the {\it sign problem}, see, for,
example,~\cite{Forcrand}, though some mathematical extensions of
lattice techniques~\cite{mathlattice} can probe this region.
Schwinger-Dyson equation techniques can also be employed
to explore all region of the phase space~\cite{Roberts}.

On the other hand a number of different model approaches indicate
that the transition along the $\mu$ axis, at $T=0$, is strongly
first order~\cite{first-order}. Since the first order line
originating at $T=0$ cannot end at the $\mu=0$ axis which
corresponds to the starting point of the cross-over line, it must
terminate somewhere in the middle of the phase diagram. This point
is generally referred to as the critical end point (CEP). The
location and observation of the CEP continue to be at the center
of efforts to understand the properties of strongly interacting
matter under extreme conditions. The mathematical extensions of
lattice techniques place the CEP in the region
$(\mu^{\mbox{\tiny{CEP}}}/T_c,T^{\mbox{\tiny{CEP}}}/T_c)\sim(1.0-1.4,0.9-0.95)$~\cite{Sharma}.

In the first of Refs.~\cite{Roberts}, it is argued that the
theoretical location of the CEP depends on the size of the
confining length scale used to describe strongly interacting
matter at finite density/temperature. This argument is supported
by the observation that the models which do not account for this
scale~\cite{NJL,pNJL,Chqm,pChqm} produce either a CEP closer to
the $\mu$ axis ($\mu^{\mbox{\tiny{CEP}}}/T_c$ and
$T^{\mbox{\tiny{CEP}}}/T_c$ larger and smaller, respectively) or a
lower $T_c$~\cite{nNJL} than the lattice based approaches or the
ones which consider a finite confining length scale. Given the
dual nature of the QCD phase transition, it is interesting to
explore whether there are other features in models which have
access only to the chiral symmetry restoration facet of QCD that,
when properly accounted for, produce the CEP's location more in
line with lattice inspired results.

An important clue is provided by the behavior of the critical
temperature as a function of an applied magnetic field. Lattice
calculations have found that this temperature decreases when the
field strength increases~\cite{Fodor,Bali:2012zg,Bali2}. It has
been recently shown that this phenomenon, dubbed {\it inverse
magnetic catalysis}, is not due exclusively to confinement but
instead that chiral symmetry restoration plays an important role.
This result is born out of the decrease of the coupling constant
with increasing field strength and is obtained within effective
models that do not have confinement such as the Abelian Higgs
model or the linear sigma model with quarks. The novel feature
implemented in these calculations is the handling of the screening
properties of the plasma, which effectively makes the treatment go
beyond the mean field approximation~\cite{Ayala1, Ayala2}.
The importance of accounting for screening in plasmas where
massless bosons appear has been pointed out  since the pioneering
work in Ref.~\cite{Jackiw} and implemented in the context of the
Standard Model to study the electroweak phase
transition~\cite{Carrington}. Screening is also important to
obtain a decrease of the coupling constant with the magnetic field
strength in QCD in the Hard Thermal Loop
approximation~\cite{Ayala3}. 

In this work we explore the
consequences of the proper handling of the plasma screening
properties in the description of the effective QCD phase diagram
within the linear sigma model with quarks. We argue that it is the adequate description of the
plasma screening properties for the chiral symmetry breaking
within the model which determines the CEP's location. Since the linear sigma model does not exhibit
confinement, one could think that the present calculation refers to finding the location of the CEP associated to chiral symmetry restoration. However, although at present there are no theoretical calculations, stemming from first principles, that prove the coincidence of chiral symmetry restoration and deconfinement transitions, there is nevertheless evidence that this may be the case, at least for not too high values of the chemical potential. The evidence is provided by lattice studies for $\mu=0$ and $T \neq 0$. Their computation of the Polyakov loop and the light quark condensate susceptibilities indicates that the transition regions coincide. Calculations based on solutions of Schwinger-Dyson equations are in agreement with this statement as well~\cite{SDCEP}. Therefore, even though strictly speaking our approach can only access chiral symmetry restoration, we emphasize that both transitions can also take place around the same region in the $(\mu, T)$ plane and therefore that, to have access to the critical values for either transition, it may be enough to only look at one of the transition aspects.
We find that for certain values of the model parameters, obtained from physical
constraints, the CEP's location agrees with lattice inspired calculations. 

We start from the linear sigma model coupled to quarks. It is
given by the Lagrangian density
\begin{eqnarray}
   {\mathcal{L}}&=&\frac{1}{2}(\partial_\mu \sigma)^2  + \frac{1}{2}(\partial_\mu \vec{\pi})^2 + \frac{a^2}{2} (\sigma^2 + \vec{\pi}^2) \nonumber \\
   &-& \frac{\lambda}{4} (\sigma^2 + \vec{\pi}^2)^2 
   + i \bar{\psi} \gamma^\mu \partial_\mu \psi \nonumber \\
   &-& g\bar{\psi} (\sigma + i \gamma_5 \vec{\tau} \cdot \vec{\pi} )\psi,
\label{lagrangian}
\end{eqnarray}
where $\psi$ is an SU(2) isospin doublet, $\vec{\pi}=(\pi_1,
\pi_2, \pi_3 )$ is an isospin triplet and $\sigma$ is an isospin
singlet. The neutral pion is taken as the third component of the
pion isovector, $\pi^0=\pi_3$ and the charged pions as
$\pi_\pm=(\pi_1\mp i\pi_2)/2$. The squared mass parameter $a^2$
and the self-coupling $\lambda$ and $g$ are taken to be positive.

To allow for the spontaneous breaking of symmetry, we let the
$\sigma$ field develop a vacuum expectation value $v$ 
\bea
   \sigma \rightarrow \sigma + v,
\label{shift} 
\eea 
which can later be taken as the order parameter
of the theory.  After this shift, the Lagrangian density can be
rewritten as 
\bea
   {\mathcal{L}} &=& -\frac{1}{2}\sigma\partial_{\mu}\partial^\mu\sigma-\frac{1}
   {2}\left(3\lambda v^{2}-a^{2} \right)\sigma^{2}\nn
   &-&\frac{1}{2}\vec{\pi}\partial_{\mu}\partial^\mu\vec{\pi}-\frac{1}{2}\left(\lambda v^{2}- a^2 \right)\vec{\pi}^{2}+\frac{a^{2}}{2}v^{2}\nn
  &-&\frac{\lambda}{4}v^{4} + i \bar{\psi} \gamma^\mu \partial_\mu \psi
  -gv \bar{\psi}\psi + {\mathcal{L}}_{I}^b + {\mathcal{L}}_{I}^f,
  \label{lagranreal}
\eea
where ${\mathcal{L}}_{I}^b$ and  ${\mathcal{L}}_{I}^f$ are given by
\begin{eqnarray}
  {\mathcal{L}}_{I}^b&=&-\frac{\lambda}{4}\Big[(\sigma^2 + (\pi^0)^2)^2\nn
  &+& 4\pi^+\pi^-(\sigma^2 + (\pi^0)^2 + \pi^+\pi^-)\Big],\nn
  {\mathcal{L}}_{I}^f&=&-g\bar{\psi} (\sigma + i \gamma_5 \vec{\tau} \cdot \vec{\pi} )\psi,
  \label{lagranint}
\end{eqnarray}
and describe the interactions among the fields $\sigma$,
$\vec{\pi}$ and $\psi$, after symmetry breaking. From
Eq.~(\ref{lagranreal}) we see that the $\sigma$, the three pions
and the quarks have masses 
\bea
  m^{2}_{\sigma}&=&3  \lambda v^{2}-a^{2},\nn
  m^{2}_{\pi}&=&\lambda v^{2}-a^{2}, \nn
  m_{f}&=& gv,
\label{masses}
\eea
respectively.

The one-loop effective potential for the linear sigma model with quarks 
including the plasma screening properties encoded in the ring diagrams contribution has been calculated in detail for zero chemical potential in previous works~\cite{Ayala:2009ji, Ayala:2014mla}. 
It is given by
\bea
V^{({\mbox{\small{eff}}})}&=&
-\frac{a^2}{2}v^2 + \frac{\lambda}{4}v^4\nn
&+&\sum_{i=\sigma,\vec{\pi}}\left\{\frac{m_i^4}{64\pi^2}
\left[ \ln\left(\frac{(4\pi T)^2}{2a^2}\right)\right.\right. \nn
&-&\left. 2\gamma_E +1\frac{}{}\right] 
- \frac{\pi^2T^4}{90} + \frac{m_i^2T^2}{24}\nn
&-&\left. \frac{T}{12 \pi}(m_i^2 + \Pi)^{3/2} \right\} \nn
&-& N_{c}\sum_{f=u,d}\left\{\frac{m_f^4}{16\pi^2}
\left[\ln\left(\frac{(\pi T)^2}{2a^2}\right)\right.\right.\nn
&-&\left.2\gamma_{\text{E}}+1\right] \left. -\frac{m_f^2T^2}{12} + \frac{7\pi^{2}T^{4}}{180} \right\},
\label{Veff-mu0}
\eea
where $\Pi=\frac{\lambda T^2}{2}+\frac{N_fN_cg^2T^2}{6}$ is the 
self-energy for any of the bosons. The analysis 
in Refs.~\cite{Ayala:2009ji, Ayala:2014mla} shows that the inclusion of the ring 
diagrams renders the effective potential stable. 

When the chemical potential is non-vanishing, the calculation of the effective
potential is more complicated. Though the boson contribution 
remains the same, the fermion contribution has to be modified due to the chemical potential. 
The modification enters the calculation in two ways: 
indirectly into the boson self-energy and 
directly from its contribution to the effective potential. We compute 
explicitly the latter below.

To one-loop order the fermion contribution to the effective potential in the
imaginary time formalism of thermal field theory is given 
by~\cite{Ayala:2009ji}
\begin{eqnarray}
V_{f}&\!\!\!\!=\!\!\!\!&-\frac{2}{\beta}\int\df{k}
\left[\beta\omega + \ln\left(1+\me^{-\beta(\omega-\mu)}\right)\right. \nn
&\!\!\!\!+\!\!\!\!& \left. \ln\left(1+\me^{-\beta(\omega+\mu)} \right) \right],
\label{Vf}
\end{eqnarray}
where $\beta=T^{-1}$ and $\omega=(\vec{k}^{2}+m_{f}^{2})^{1/2}$, and
the sum over the fermion Matsubara frequencies has been performed. The first term in Eq.~(\ref{Vf}) corresponds to the vacuum contribution whereas the second and third ones are the matter contributions. Note that the matter contribution is made out of separate quark and antiquark pieces due to the finite
chemical potential. The vacuum contribution is well-known~\cite{Ayala:2009ji} and can be expressed, after mass renormalization as a function of the renormalization scale $\tilde{\mu}$. For the evaluation of the medium's contribution in Eq.~(\ref{Vf}) we will adapt the technique from Ref.~\cite{Jackiw} to the 
present case.

The main idea of this method is to produce a
second-order differential equation in $y^{2}$, where $y=m_{f}/T$, valid at high temperature with
$m_{f}$ as the smallest of all scales, for the finite temperature part of the 
potential, which we will denote by $\tilde{V}_{f}$, given in 
Eq.~(\ref{Vf}) with appropriate boundary conditions at $y=0$, where the integrals can be analytically evaluated. The expression for the effective potential is obtained by 
integrating this differential equation and using the given boundary conditions. 
The second-order differential equation satisfied by $\tilde{V}_{f}$ is
\begin{eqnarray}
\frac{d^{2}\tilde{V}_{f}}{dy^{4}}&\!\!\!\!=\!\!\!\!&\frac{1}{8\pi^{2}\beta^{4}}
\left[\ln\left(\frac{y^{2}}{(4\pi)^{2}}\right)
-\psi^{0}\left(\frac{1}{2}+\frac{\mi z}{2\pi}\right) \right. \nn
&\!\!\!\!-\!\!\!\!& \left.\psi^{0}\left(\frac{1}{2}-\frac{\mi z}{2\pi}\right)\right],
\label{diffeqn}
\end{eqnarray}
where $\psi^0(x)$ is the digamma function. The boundary conditions 
are
\begin{eqnarray}
\tilde{V}_{f}|_{y^{2}=0}&\!\!\!\!=\!\!\!\!&
\frac{2}{\pi^{2}\beta^{4}}\left[Li_4(-e^{z})+Li_4(-e^{-z})\right]\nn
\frac{d\tilde{V}_{f}}{dy^{2}}|_{y^{2}=0}&\!\!\!\!=\!\!\!\!&
\frac{-1}{2\pi^{2}\beta^{4}}\left[Li_2(-e^{z})+Li_2(-e^{-z})\right],\nn
\label{boundarycond}
\end{eqnarray}
where $Li_n(x)$ is the polylogarithm function of order $n$. 

The boundary conditions Eqs.~(\ref{boundarycond}) fix the two 
integration constants needed to determine $\tilde{V}_{f}(y,z)$.
The solution of Eq.~(\ref{diffeqn}) that satisfies the boundary conditions
Eqs.~(\ref{boundarycond}) is given by
\begin{eqnarray}
\tilde{V}_{f}&\!\!\!\!=\!\!\!\!&-\frac{1}{16\pi^{2}\beta^{4}}
\left\{-y^{4}\ln\left(\frac{y^{4}}{(4\pi)^{2}}\right)
\right. \nn
&\!\!\!\!+\!\!\!\!& y^{4}\left[\frac{3}{2} + \psi^{0}\left(\frac{1}{2}+\frac{\mi z}{2\pi}\right)
+\psi^{0}\left(\frac{1}{2}-\frac{\mi z}{2\pi}\right)\right] \nn
&\!\!\!\!+\!\!\!\!& 8y^{2}\left[Li_2(-e^{z})+Li_2(-e^{-z})\right] \nn
&\!\!\!\!-\!\!\!\!& \left. 32\left[Li_4(-e^{z})+Li_4(-e^{-z})\right]\right\}.
\label{Vfexplpre}
\end{eqnarray}
Combining the vacuum contribution after mass renormalization with the 
finite temperature part and recalling that $y=m_{f}/T$ and $z=\mu/T$ we 
finally have
\begin{eqnarray}
\tilde{V}_{f}&\!\!\!\!=\!\!\!\!&-\frac{1}{16\pi^{2}}
\left\{ m_{f}^{4}\left[\ln\left(\frac{(4\pi T)^{2}}{2\tilde{\mu}^{2}}\right) 
\right.\right. \nn
&\!\!\!\!+\!\!\!\!& \left. \psi^{0}\left(\frac{1}{2}+\frac{\mi\mu }{2\pi T}\right)
+ \psi^{0}\left(\frac{1}{2}-\frac{\mi\mu}{2\pi T}\right)\right] \nn
&\!\!\!\!+\!\!\!\!& 8m^{2}T^{2}\left[Li_2(-e^{\mu/T})+Li_2(-e^{-\mu/T})\right] \nn
&\!\!\!\!-\!\!\!\!& \left. 32T^{4}\left[Li_4(-e^{\mu/T})+Li_4(-e^{-\mu/T})\right]\right\}. \nn
\label{Vfexpl}
\end{eqnarray}

It can also be shown that the boson self-energy $\Pi$ 
computed for a finite chemical potential and in the limit where the 
masses are small compared to $T$, is given by
\bea
\Pi&\!\!\!\!=\!\!\!\!&\frac{\lambda T^2}{2} \nonumber \\
&\!\!\!\!-\!\!\!\!&\frac{N_fN_cg^2T^2}{\pi^2}\nn
&\!\!\!\!\times\!\!\!\!&\left[ Li_2(-e^{\mu/T}) + Li_2(-e^{-\mu/T})\right]
\label{self-energy} 
\eea 
where $N_f=2$ and $N_c=3$ are the number of light flavors and colors,
respectively.


Choosing the renormalization scale as $\tilde{\mu}=e^{-1/2}a$, the effective 
potential up to the ring diagrams contribution for a finite chemical potential 
and in the limit where the masses are small compared to $T$, is then given by

\bea
   \!\!\!\!\!\!\!\!V^{({\mbox{\small{eff}}})}&\!\!\!\!=\!\!\!\!&
   -\frac{a^2}{2}v^2 + \frac{\lambda}{4}v^4\nn
  \!\!\!\! \!\!\!\!&\!\!\!\!+\!\!\!\!&\sum_{i=\sigma,\vec{\pi}}\left\{\frac{m_i^4}{64\pi^2}
   \left[ \ln\left(\frac{(4\pi T)^2}{2a^2}\right)
   \right.\right.\nn
   \!\!\!\!\!\!\!\!&\!\!\!\!-\!\!\!\!&\left. 2\gamma_E +1\frac{}{}\right]-\frac{\pi^2T^4}{90} + \frac{m_i^2T^2}{24}\nn
   \!\!\!\!\!\!\!\!&\!\!\!\!-\!\!\!\!& \left.\frac{T}{12 \pi}(m_i^2 + \Pi)^{3/2} \right\} \nn
   \!\!\!\!\!\!\!\!&\!\!\!\!-\!\!\!\!& \frac{N_{c}}{16\pi^2}\sum_{f=u,d}
   \left\{m_f^4\left[\ln\left(\frac{(4\pi T)^2}{2a^2}\right)+1\right.\right.\nn
   \!\!\!\!\!\!\!\!&\!\!\!\!+\!\!\!\!&\left.\psi^0\left(\frac{1}{2}+\frac{i\mu}{2\pi T}\right) +
          \psi^0\left(\frac{1}{2}-\frac{i\mu}{2\pi T}\right) \right] \nn
   \!\!\!\!\!\!\!\!&\!\!\!\!+\!\!\!\!&8\ m_f^2T^2\left[ Li_2(-e^{\mu/T}) + Li_2(-e^{-\mu/T})\right]\nn
   \!\!\!\!\!\!\!\!&\!\!\!\!-\!\!\!\!& \left. 32\ T^4 \left[ Li_4(-e^{\mu/T}) + Li_4(-e^{-\mu/T}) \right]\right\}. 
\nn
\label{Veff-mid}
\eea

In the limit when $\mu\rightarrow 0$, Eq.~(\ref{Veff-mid}) becomes Eq.~(\ref{Veff-mu0}). In the same limit,
Eq.~(\ref{self-energy}) reduces to the well known expression for the self-energy at
high temperature~\cite{Ayala2}.

Note that the self-energy provides the screening to render the
effective potential in Eq.~(\ref{Veff-mid}) stable. Should this
self-energy be absent, the term $(m_i^2 + \Pi)^{3/2}$ would
instead be $(m_i^2)^{3/2}$, which becomes imaginary when for
certain values of $v$,  $m_i^2$ becomes negative [see
Eqs.~(\ref{masses})]. This term is obtained from considering the
resummation of the ring diagrams and therefore
Eq.~(\ref{Veff-mid}) represents the effective potential computed
beyond the mean field approximation that accounts for the leading
screening effects at high temperature.

\begin{figure}[t!]
\begin{center}
\includegraphics[scale=0.85]{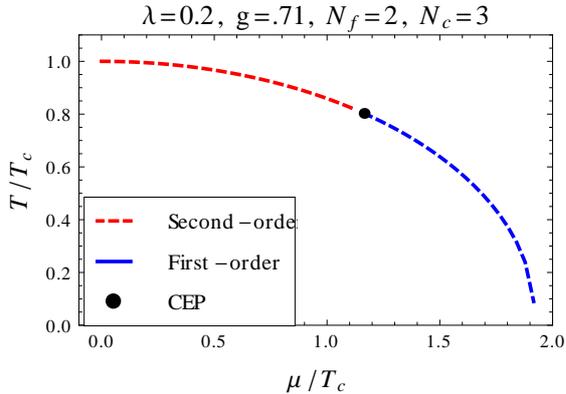}
\end{center}
\caption{Effective QCD phase diagram computed for $\lambda = 0.2$ and $g = 0.71$ obtained by considering $m_\sigma^{\mbox{\tiny{vac}}} = 300$ MeV. For small values of $\mu$
the phase transition is second order. The order of the transition
changes to first order for larger values of $\mu$. The CEP is
located at
$(\mu^{\mbox{\tiny{CEP}}}/T_c,T^{\mbox{\tiny{CEP}}}/T_c)\sim(1.2,0.8)$.}
\label{fig1}
\end{figure}

In order to find the values of the parameters $\lambda$, $g$ and $a$ appropriate for the description of the phase transition, we note that when considering the thermal effects the boson masses are modified since they acquire a thermal component. For $\mu=0$ they become
\bea
   m_\sigma^2(T)&=&3\lambda v^2 -a^2 + \frac{\lambda T^2}{2}+\frac{N_fN_cg^2T^2}{6}\nn
   m_\pi^2(T)&=&\lambda v^2 -a^2 + \frac{\lambda T^2}{2}+\frac{N_fN_cg^2T^2}{6}.\nonumber\\
\label{massmod} 
\eea 
At the phase transition, the curvature of the
effective potential vanishes for $v=0$. Since the boson thermal
masses are proportional to this curvature, these also vanish at
$v=0$. From any of the Eqs.~(\ref{massmod}), we obtain a relation
between the model parameters at $T_c$ given by
\bea
   a=T_c\sqrt{\frac{\lambda}{2}+\frac{N_fN_cg^2}{6}}.
\label{relation}
\eea
Furthermore, we can fix the value of $a$ by noting from Eqs.~(\ref{masses}) that the vacuum boson masses satisfy
\bea
   a=\sqrt{\frac{m_\sigma^2 - 3m_\pi^2}{2}}.
\label{massvac}
\eea
\begin{figure}[t!]
\begin{center}
\includegraphics[scale=0.85]{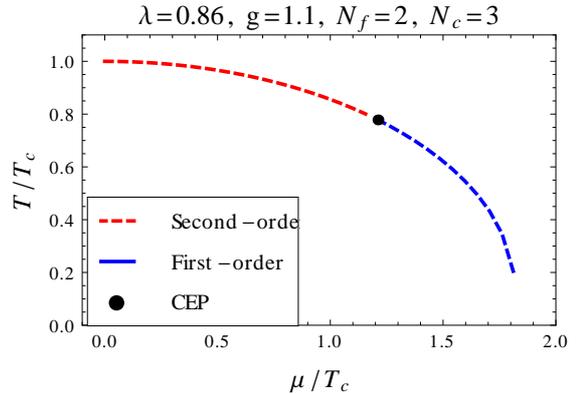}
\end{center}
\caption{Effective QCD phase diagram computed for $\lambda = 0.86$ and $g =1.1$ obtained by considering $m_\sigma^{\mbox{\tiny{vac}}} = 400$ MeV. For small values of $\mu$
the phase transition is second order. The order of the transition
changes to first order for larger values of $\mu$. The CEP is
located at
$(\mu^{\mbox{\tiny{CEP}}}/T_c,T^{\mbox{\tiny{CEP}}}/T_c)\sim(1.2,0.8)$.}
\label{fig2}
\end{figure}
Since in our scheme we consider two-flavor massless quarks in the chiral limit, we take $T_c\simeq 170$ MeV~\cite{example} which is slightly larger than $T_c$ obtained in $N_f=2+1$ lattice simulations. Also, in order to allow for a crossover phase transition for $\mu=0$ (which in our description corresponds to a second order transition) with $g$, $\lambda \sim 1$ we need that $g^2 > \lambda$. To justify the perturbative expansion we need to look for coupling constant values not too large. Furthermore since the effective potential is written as an expansion in powers of $a/T$ we need that this ratio is not too much larger than 1 (there are numerical factors in Eq.(\ref{Veff-mid}) that make it possible to consider values for $a/T_c$ slightly larger than 1). From Eqs.~(\ref{relation}) and~(\ref{massvac}) the coupling constants are proportional to $m_\sigma$ which, from the above conditions, restricts the analysis to considering not too large values of $m_\sigma$. Since the purpose of this work is not to pursue a precise determination of the couplings but instead to call attention to the fact that the proper treatment of screening effects allows the linear sigma model to provide solutions for the CEP, we consider small values for $m_\sigma$. The Particle Data Group quotes 400 MeV $\leq m_\sigma \leq$ 550 MeV~\cite{PDG}. There are also analyses that place $m_\sigma$ close to the two-pion threshold~\cite{lowmass}. Given that $\sigma$ is anyhow a broad resonance, in order to satisfy the above requirements let us first take for definitiveness two values $m_\sigma = 300$ and $400$ MeV, namely, close to the two-pion threshold. Therefore, the allowed values for the couplings $\lambda$ and $g$ are restricted by
\bea
   \sqrt{\frac{\lambda}{2}+\frac{N_fN_cg^2}{6}} = 0.77\ , 1.28.
\label{allowed}
\eea
Equation~(\ref{allowed}) provides a relation between $\lambda$ and $g$. A possible solution consistent with the above requirements is given by $\lambda=0.2$, $g=0.71$ for $m_\sigma = 300$ MeV and by $\lambda=0.86$, $g=1.1$ for $m_\sigma = 400$ MeV.

Figures~\ref{fig1} and~\ref{fig2} show the phase diagram obtained for the sets of
allowed values $\lambda$ and $g$ for $m_\sigma = 300, 400$ MeV. Note that for small $\mu$ the
phase transition is second order. In this case the
(pseudo)critical temperature is determined from setting the second
derivative of the effective potential in Eq.~(\ref{Veff-mid}) to
zero at $v=0$. When $\mu$ increases, the phase transition becomes
first order. The critical temperature is now computed by looking
for the temperature where a secondary minimum for $v\neq 0$ is
degenerate with a minimum at $v=0$. In both of these cases, from the detailed analysis, we
locate the position of the CEP as
$(\mu^{\mbox{\tiny{CEP}}}/T_c,T^{\mbox{\tiny{CEP}}}/T_c)\sim(1.2,0.8)$,
which is in the same range as the CEP found from lattice inspired
analyses~\cite{mathlattice}. Note also that the phase transition curve is essentially flat
close to the $T$ axis which goes also in line with the results of Ref.~\cite{curvature}. 

\begin{figure}[t!]
\begin{center}
\includegraphics[scale=0.85]{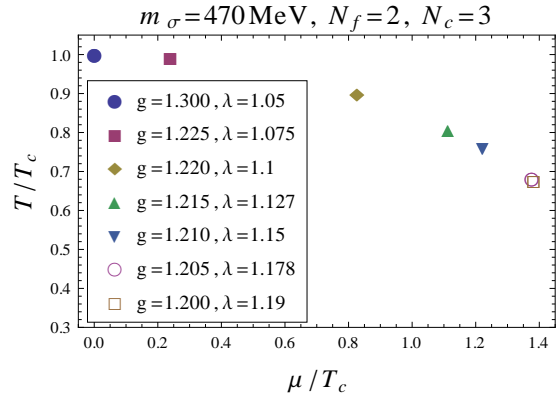}
\end{center}
\caption{CEP's location for $m_\sigma = 470$ MeV for different values of $\lambda$ and $g$, with the condition $g>\lambda$ and consistent with Eq.~(\ref{relation}).}
\label{fig3}
\end{figure}

Figure~\ref{fig3} shows the location of the CEP for $m_\sigma = 470$ MeV (which is at the upper edge for the values allowed in our approach) when varying $\lambda$ and $g$ according to Eq.~(\ref{relation}) with the condition $g>\lambda$. We observe that a small difference between $\lambda$ and $g$ favors a CEP's location away from the vertical axis and that when the difference between the couplings increases, the CEP moves toward the $T$-axis. We can understand this behavior by noticing that since $g$ controls the strength of the fermion's contribution to the effective potential and that the first order nature of the transition is governed by these particles, when $\lambda$ and $g$ are similar, the first order phase transition starts for larger values of $\mu$. When this difference increases, the fermion contribution to the effective potential is important to start with and thus the first order transitions start for smaller values of $\mu$. Also note that although the CEP's location shows a large dispersion, we still find solutions consistent with the location obtained for the other two explored values of $m_\sigma$.

We thus see that the allowed values for the couplings are not uniquely determined. In order to discriminate between different sets of couplings and further constrain their values, one could resort to study another observable such as the pressure. As shown by lattice QCD~\cite{pressure}, this quantity is a monotonically increasing function of temperature. We have performed preliminary studies that show that there are sets, among the explored ones, that satisfy this behavior for the pressure. A more detailed analysis focusing on this observable is on its way and will be reported elsewhere. Nevertheless, we point out that in order to have access to a more precise determination of the couplings, the model needs first to be refined to either include higher powers of $a/T$ in the analysis or to allow a description for the case with $a/T > 1$.

In conclusion, we have shown that it is possible to obtain values for the couplings that allow to locate the CEP in the region found by mathematical extensions
of lattice analyses. Since the linear sigma model does not 
have confinement we attribute this location to the adequate description
of the plasma screening properties for the chiral symmetry
breaking at finite temperature and density. These properties are
included into the calculation of the effective potential through
the boson's self-energy and in the determination of the allowed
range for the coupling constants through the observation that the
thermal boson masses vanish at the phase transition for $\mu=0$. 
These observations determine a relation between the model parameters which is 
put in quantitative terms by taking physical values for $T_c$ from
lattice calculations and for $a$ from the vacuum boson masses. We believe this description will
play an important role in determining the location of the CEP also in
QCD.

\section*{Acknowledgments}

A. A. acknowledges useful comments from M. E. Tejeda-Yeomans,
M. Loewe and A. J. Mizher. J. J. Cobos-Mart\1nez acknowledges support from a CONACyT-Mexico postdoctoral fellowship under contract number 290917-UMSNH. Support for this work has
been received in part from UNAM-DGAPA-PAPIIT grant number 101515, from CONACyT-M\'exico grant number 128534 and from CIC-UMSNH grant numbers 4.10 and 4.22.

\end{document}